\documentclass[aps,pra,twocolumn,showpacs,floatfix]{revtex4} 
\usepackage{amsmath,amssymb,isolatin1,graphicx,times}

\newcommand{\be}{\begin{equation}}
\newcommand{\ee}{\end{equation}}
\newcommand{\bea}{\begin{eqnarray}}
\newcommand{\eea}{\end{eqnarray}}
\newcommand{\bw}{\begin{widetext}}
\newcommand{\ew}{\end{widetext}}

\newcommand{\rd}{{\rm d}}

\newcommand{\kommentar}[1]{}

\begin{document}
 
\title{Spacetime structures of continuous time quantum walks}
\author{Oliver M{\"u}lken}
\author{Alexander Blumen}
\affiliation{
Theoretische Polymerphysik, Universit\"at Freiburg,
Hermann-Herder-Straße 3, 79104 Freiburg i.Br., Germany}

\date{\today} 
\begin{abstract}
The propagation by continuous time quantum walks (CTQWs) on
one-dimensional lattices shows structures in the transition probabilities
between different sites reminiscent of quantum carpets.  For a system with
periodic boundary conditions, we calculate the transition probabilities
for a CTQW by diagonalizing the transfer matrix and by a Bloch function
ansatz.  Remarkably, the results obtained for the Bloch function ansatz
can be related to results from (discrete) generalized coined quantum
walks. Furthermore, we show that here the first revival time turns out to
be larger than for quantum carpets.
\end{abstract}
\pacs{05.60.Gg,05.40.-a,03.67.-a}
\maketitle

Simple theoretical models have always been very useful for our
understanding of physics.  In quantum mechanics, next to the harmonic
oscillator, the particle in a box provides much insight into the quantum
world (e.g.\ \cite{Sakurai}). Recently, the problem of a quantum
mechanical particle initially characterized by a gaussian wave packet and
moving in an infinite box has been reexamined \cite{kinzel1995,
fgrossmann1997, berry1996}. Surprisingly, this simple system shows complex
but regular spacetime probability structures which are now called quantum
carpets.

In solid state physics and quantum information theory, one of the most
simple systems is associated with a particle moving in a regular periodic
potential. This can be, for instance, either an electron moving through a
crystal \cite{Ziman, Kittel} or a qubit on an optical lattice or in an
optical cavity \cite{kempe2003,knight2003,duer2002}.  For the electron
moving through a crystal, the band structure and eigenfunctions are well
known. In principle, the same holds for the qubit.  However, in quantum
information theory, the qubit on a lattice or, more general, on a graph is
used to define the quantum analog of a random walk. As classically, there
is a discrete \cite{aharonov1993} and a continuous-time \cite{farhi1998}
version. Unlike in classical physics, these two are not translatable into
each other. 

Here we focus on continuous-time (quantum) random walks.  Consider a walk
on a graph which is a collection of connected nodes.  Lattices are very
simple graphs where the nodes are connected in a very regular manner. To
every graph there exists a corresponding adjacency or connectivity matrix
${\bf A} = (A_{ij})$, which is a discrete version of the Laplace operator.
The non-diagonal elements $A_{ij}$ equal $-1$ if nodes $i$ and $j$ are
connected by a bond and $0$ otherwise. The diagonal elements $A_{ii}$
equal the number of bonds which exit from node $i$, i.e., $A_{ii}$ equals
the functionality $f_i$ of the node $i$. 

Classically, a continuous-time random walk (CTRW) is governed by the
master equation \cite{weiss,vankampen}
\be
\frac{\rm d}{{\rm d} t} p_{jk}(t) = \sum_l T_{jl} \ p_{lk}(t),
\label{mast_eq0}
\ee
where $p_{jk}(t)$ is the conditional probability to find the CTRW at time
$t$ at node $j$ when starting at node $k$. The transfer matrix of the
walk, ${\bf T} = (T_{jk})$, is related to the adjacency matrix by ${\bf T}
= - \gamma {\bf A}$, where we assume the transmission rate $\gamma$ of all
bonds to be equal for simplicity. Formally, this approach can be
generalized to continuous models like the Lorentz gas \cite{mvb2004}.  The
formal solution of Eq.(\ref{mast_eq0}) is
\be
p_{jk}(t) = \langle j | e^{{\bf T} t} | k \rangle.
\label{cl_prob}
\ee

The quantum-mechanical extension of a CTRW is called continuous-time
quantum walk (CTQW). These are obtained by identifying the Hamiltonian of
the system with the (classical) transfer operator, ${\bf H} = - {\bf T}$
\cite{farhi1998,childs2002,mb2004a}. Then the basis vectors $|k\rangle$
associated with the nodes $k$ of the graph span the whole accessible
Hilbert space.  In this basis the Schr\"odinger equation (SE) reads
\be
i \frac{\rm d}{{\rm d} t} | k \rangle = {\bf H} | k \rangle,
\label{sgl}
\ee
where we have set $m\equiv1$ and $\hbar\equiv1$. The time evolution of a
state $| k \rangle$ starting at time $t_0$ is given by $| k(t) \rangle =
{\bf U}(t,t_0) | k \rangle$, where ${\bf U}(t,t_0) = \exp(-i {\bf H}
(t-t_0))$ is the quantum mechanical time evolution operator. Now the
transition amplitude $\alpha_{jk}(t)$ from state $| k \rangle$ at time $0$
to state $|j\rangle$ at time $t$ reads
\be
\alpha_{jk}(t) = \langle j | e^{-i {\bf H} t} | k \rangle.
\label{qm_ampl}
\ee
Following from Eq.(\ref{sgl}) the $\alpha_{jk}(t)$ obey
\be
i \frac{\rm d}{{\rm d} t} \alpha_{jk}(t) = \sum_l H_{jl}
\alpha_{lk}(t).
\label{sgl_ampl}
\ee
The main difference between Eq.(\ref{cl_prob}) and Eq.(\ref{qm_ampl}) is
that classically $\sum_j p_{jk}(t) = 1$, whereas quantum mechanically
$\sum_j |\alpha_{jk}(t)|^2 =1$ holds.

In principle, for the full solution of Eqs.(\ref{mast_eq0}) and
(\ref{sgl_ampl}) all the eigenvalues {\sl and} all the eigenvectors of
${\bf T}=-{\bf H}$ (or, equivalently, of ${\bf A}$) are needed.  Let
$\lambda_n$ denote the $n$th eigenvalue of ${\bf A}$ and ${\bf \Lambda}$
the corresponding eigenvalue matrix. Furthermore, let ${\bf Q}$ denote the
matrix constructed from the orthonormalized eigenvectors of ${\bf A}$, so
that ${\bf A} = {\bf Q}{\bf\Lambda}{\bf Q}^{-1}$. Now the classical
probability is given by
\be
p_{jk}(t) = \langle j| {\bf Q} e^{-t \gamma{\bf \Lambda}} {\bf Q}^{-1} | k
\rangle,
\label{cl_prob_full}
\ee
whereas the quantum mechanical transition probability is
\be
\pi_{jk}(t) \equiv |\alpha_{jk}(t)|^2 = |\langle j| {\bf Q} e^{- i t
\gamma {\bf
\Lambda}} {\bf Q}^{-1} | k \rangle|^2.
\label{qm_prob_full}
\ee

The unitary time evolution prevents that $\pi_{jk}(t)$ has a definite
limit for $t\to\infty$. In order to compare the classical long time
probability with the quantum mechanical one, one usually uses the limiting
probability distribution \cite{aharonov2001}
\be
\chi_{kj} \equiv \lim_{T\to\infty} \frac{1}{T} \int_0^T \rd t \ \pi_{jk}(t).
\ee

In the subsequent calculation we restrict ourselves to CTQWs on regular
one-dimensional (1d) lattices. Then the adjacency matrix ${\bf A}$ takes
on a very simple form.  For a 1d lattice with periodic boundary
conditions, i.e.\ a circle, every node has exactly two neighbors. Thus,
for a lattice of length $N$, with the boundary condition that node $N+1$
is equivalent to node $1$, we have
\be
{\bf A} = (A_{ij}) = \begin{cases} 2 & i=j \\ -1 & i= j\pm 1 \\ 0 &
\mbox{else}. \end{cases}
\label{mat1d}
\ee
For a lattice with reflecting boundary conditions the adjacency matrix
${\bf A}$ is analogous to Eq.(\ref{mat1d}), except that $A_{11} = A_{NN} =
1$ and $A_{1N}=A_{N1} = 0$ because the end nodes have only one neighbor.
Solving the eigenvalue problem for ${\bf A}$, which is a real and
symmetric matrix is a well-known problem, also of much interest in polymer
physics \cite{blumen2003,blumen2004}. A different ansatz describing the
dynamics of a quantum particle in 1d was given by W{\'o}jcik and Dorfman
who employ a quantum multibaker map \cite{wojcik2003}.

The structure of ${\bf H}=\gamma{\bf A}$ suggests an analytic treatment.
For a 1d lattice with periodic boundary conditions and $\gamma=1$ the
Hamiltonian acting on a state $|j\rangle$ is given by
\be
{\bf H} |j\rangle = 2 |j\rangle - |j-1\rangle - |j+1\rangle,
\label{hamil}
\ee
which is the discrete version of the Laplacian $-\Delta = -\nabla^2$.
Eq.(\ref{hamil}) is the discrete version of the Hamiltonian for a free
particle moving on a lattice.  It is well known in solid state physics
that the solutions of the SE for a particle moving freely in a regular
potential are Bloch functions \cite{Kittel,Ziman}.  Thus, 
the time independent SE is given by
\be
{\bf H} |\Phi_\theta\rangle = E_\theta |\Phi_\theta\rangle,
\label{tiseq}
\ee
where the 
eigenstates $|\Phi_\theta\rangle$ are Bloch states and can be written as a
linear combination of states $|j\rangle$ localized at nodes $j$,
\be
|\Phi_\theta\rangle = \frac{1}{\sqrt N}\sum_{j=1}^N e^{-i \theta j}
|j\rangle.
\label{blochef}
\ee
The projection on the state $|j\rangle$ than reads $\Phi_\theta(j) \equiv
\langle j | \Phi_\theta\rangle = e^{-i\theta j}/\sqrt{N}$, which is
nothing but the Bloch relation $\Phi_\theta(j+1) = e^{-i\theta}\Phi_\theta(j)$
\cite{Kittel,Ziman}.
Now the energy is obtained from Eqs.(\ref{tiseq}) and
(\ref{blochef}) as
\be
E_\theta = 2 - 2 \cos\theta.
\ee
For small $\theta$ the energy is given by $E_\theta \approx \theta^2$
which resembles the energy spectrum of a free particle.

With this ansatz we calculate the transition amplitudes $\alpha_{kj}(t)$.
The state $|j\rangle$ is localized at node $j$ and may be described by a
Wannier function \cite{Kittel,Ziman}, 
i.e.\ by inverting Eq.(\ref{blochef}),
\be
|j\rangle 
= \frac{1}{\sqrt N} \sum_\theta e^{i\theta j} | \Phi_\theta \rangle.
\ee
Since the states $|j\rangle$ span the whole accessible Hilbert space, we
have $\langle k | j \rangle = \delta_{kj}$ and therefore via
Eq.(\ref{blochef}) also $\langle \Phi_{\theta'} | \Phi_\theta \rangle =
\delta_{\theta'\theta}$.  

Then the transition amplitude reads
\bea
\alpha_{kj}(t) &=& 
\frac{1}{N} \sum_{\theta,\theta'} \langle \Phi_{\theta'} | e^{-i\theta k}
e^{-i{\bf H}t} e^{i\theta' j} | \Phi_\theta \rangle \nonumber \\
&=& \frac{1}{N} \sum_{\theta} e^{-iE_\theta t} e^{-i\theta(k-j)}.
\label{ampl_bloch}
\eea

The periodic boundary condition for a 1d lattice of size $N$ requires
$\Phi_\theta(N+1) = \Phi_\theta(1)$, thus $\theta = 2 n \pi /N$ with
$n\in]0,N]$. Now Eq.(\ref{ampl_bloch}) is given by
\be
\alpha_{jk}(t) = \frac{e^{-i2t}}{N} \sum_{n} e^{i2t\cos(2n\pi/N)}
e^{-i2\pi n(k-j)/N}.
\label{ampl_bloch2}
\ee
For small $\theta$, this result is directly related to the results
obtained for a quantum particle in a box \cite{kinzel1995, fgrossmann1997,
berry1996}, because then we have $E_n \sim n^2$.

In the limit $N\to\infty$, Eq.(\ref{ampl_bloch2}) translates to
\bea
\lim_{N\to\infty} \alpha_{kj}(t) &=& \frac{e^{-i2t}}{2\pi}
\int\limits_{-\pi}^{\pi} \rd \theta \ e^{-i\theta(k-j)} e^{i 2t\cos\theta
} \nonumber \\
&=& i^{k-j} e^{-i2t} J_{k-j} (2t),
\label{alpha_lim}
\eea
where $J_k(x)$ is the Bessel function of the first kind \cite{Ito}. The
same result has also been obtained with a functional integral ansatz
\cite{farhi1992}. From Eq.(\ref{alpha_lim}) we also see that the first
maxima of the transition probabilities are related to the maxima of the
Bessel function, since we have $\lim_{N\to\infty} \pi_{kj}(t) =
[J_{k-j}(2t)]^2$.  However, for an infinite lattice there is no
interference due to either backscattering at reflecting boundaries or
transmission by periodic boundaries.

For higher dimensional lattices the calculation is analogous.  
We note that the assumption of periodic boundary conditions is strictly
valid only in the limit of very large lattice sizes where the exact form
of the boundary does not matter \cite{Kittel,Ziman}.

\begin{figure}[ht]
\centerline{\includegraphics[width=0.9\columnwidth]{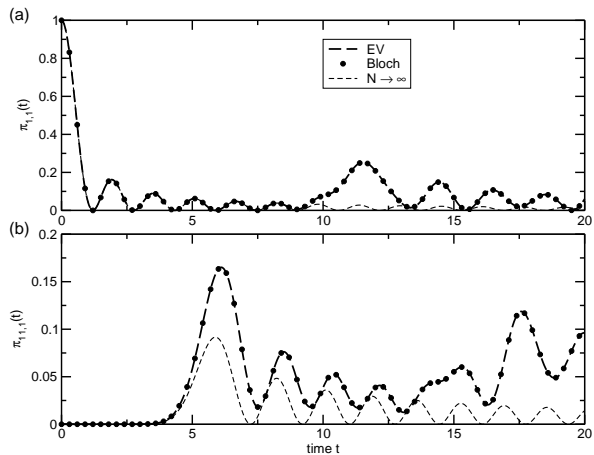}}
\caption{Plot for a CTQW on a circle of length $N=21$ of (a) the return
probability and (b) the transition probability to go in time $t$ from the
starting node to the opposite node on the circle. Time is given in units
of the inverse transmission rate $\gamma^{-1}$. The results using
Eq.(\ref{qm_prob_full}), long dashed line, and Eq.(\ref{ampl_bloch2}),
full circles, are compared to the limit $N\to\infty$, short dashed line.
}
\label{1d-circle}
\end{figure}

Very recently it has been found by W\'ojcik {\it et al.},
\cite{wojcik2004}, that the return probability for a 1d generalized coined
quantum walk (GCQW), which is a variant of a discrete quantum walk, has
the functional form $p_{kk}(t\tau) = [J_{0}(2t\sqrt{D})]^2$, where $\tau$
and $D$ are variables specified in \cite{wojcik2004}, which indeed is of
the same form as the return probability calculated from
Eq.(\ref{alpha_lim}). We interpret this as an indication that CTQWs and
GCQWs, although not directly translatable into each other, can lead to
similar results.  However, in \cite{wojcik2004} the return probability is
calculated for a particle on a very large circle such that interference
effects are not seen on the short time scales considered there. By looking
ahead at Fig.\ \ref{1d-circle}, we see that, indeed, on short time scales
this is also approximately true in our case of the CTQW on the finite
lattice. Nevertheless, without going into further detail at this point, we
note this remarkable similarity between CTQWs and GCQWs. 

For a CTQW on a 1d circular lattice we calculate the quantum mechanical
transition probabilities $\pi_{jk}(t)$.  Figure~\ref{1d-circle}(a) shows
the return probability $\pi_{kk}(t)$ for a CTQW on a circle of $21$ nodes
first evaluated in a straightforward way by diagonalizing the matrix ${\bf
A}$ numerically, then by using the Bloch function ansatz described above.
Both results coincide. For comparison we also have computed the return
probability for the infinitely extended system, see Eq.(\ref{alpha_lim}).
On small time scales all the results coincide.  At later times waves
propagating on the finite lattice start to interfere; then the results
diverge and for a finite lattice one observes an increase in the probability
of being at the starting node.  This happens around the time $t\approx
N/2$. 

In Fig.~\ref{1d-circle}(b) the probability to go from a starting node to
the farthest node on the circle, here to go from node $1$ to node $11$ (or
$12$), is plotted. Again the calculations by the eigenvalue method and by
the Bloch function ansatz are indistinguishable. As before, also the
probabilities for the infinite and for the finite systems differ. The
difference is more pronounced because in time $t\approx N/4$
counterpropagating waves from the starting node interfere at the opposite
node. 

\begin{figure}[ht]
\centerline{
\includegraphics[width=0.475\columnwidth]{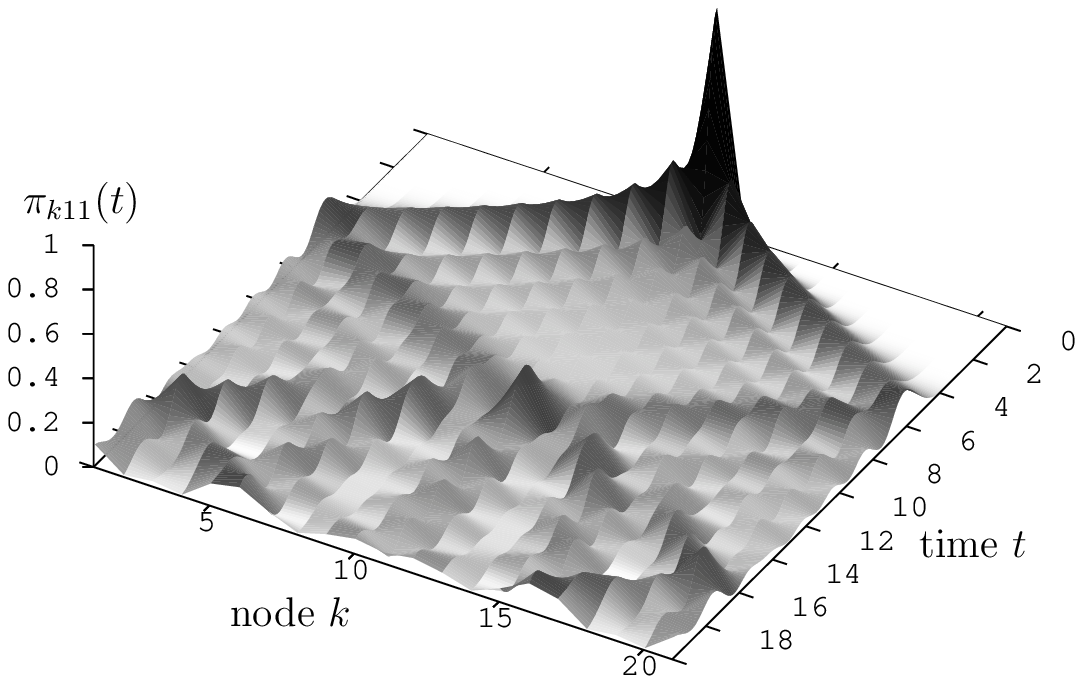}
\includegraphics[width=0.475\columnwidth]{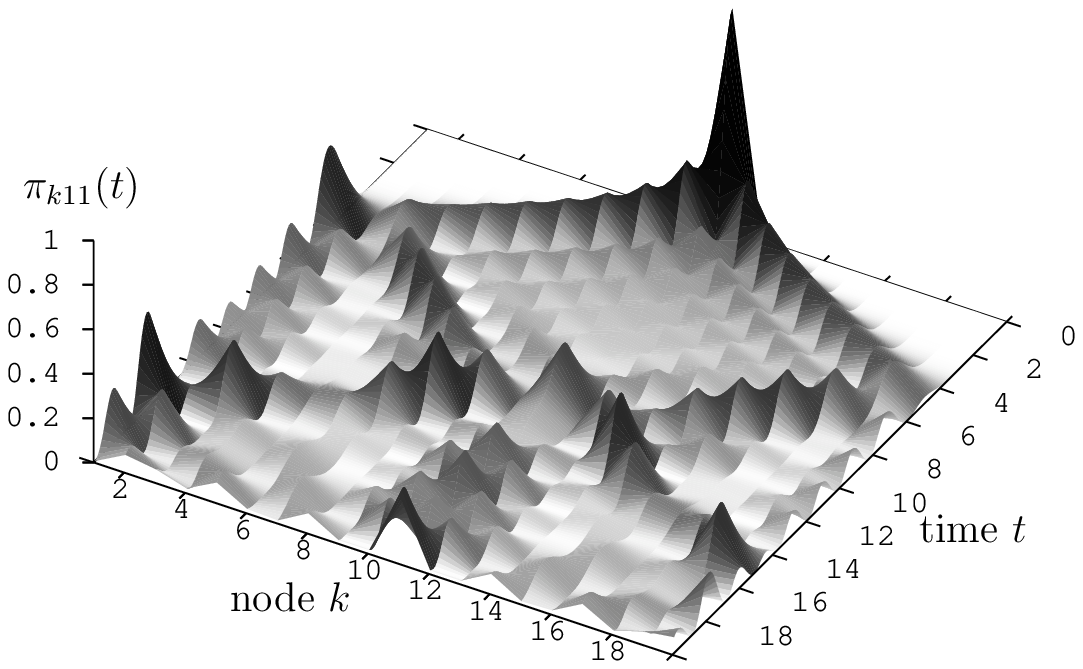}}
\caption{Plot of the probability for a CTQW on a circle of length $N=21$
(left) and $N=20$ (right) over time $t$ to go from a starting node to all
other nodes. See Fig.\ref{1d-circle} for units.
}
\label{1d-circle-cont}
\end{figure}

The probabilities to go from a starting node to all other nodes in time
$t$ on a circle of length $N=21$ is plotted in
Fig.~\ref{1d-circle-cont}(left).  (For a CTQW on a circle the starting
node is arbitrary.) For small times, when there is no interference, the
waves propagate freely.  After a time $t\approx N/4$ the waves interfere
but the pattern remains quite regular. The same holds for $N=20$, but the
structures are more regular, see Fig.~\ref{1d-circle-cont}(right). This is
due to the fact that the number of steps to go form one node to another is
even or odd in both directions for the even-numbered circle, where it is
even in one and odd in the other direction for the odd-numbered circle. 

\begin{figure}[ht]
\centerline{
\includegraphics[width=0.9\columnwidth]{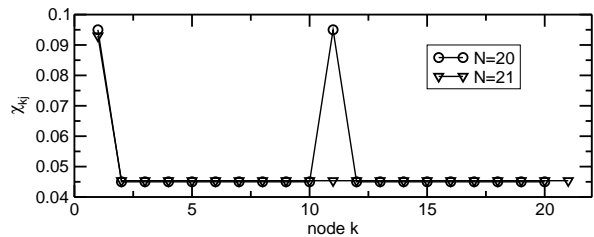}}
\caption{Limiting probability distribution $\chi_{kj}$ for a CTQW on a
circle of length $N=20$ (circles) and $N=21$ (triangles).
}
\label{1d-circle_chi}
\end{figure}
Figure \ref{1d-circle_chi} supports this. The limiting distribution
$\chi_{kj}$ has two maxima for $N=20$, one at the starting node $1$ and
one at the opposite node $11$, reflecting the higher symmetry of the
lattice. Whereas there is only one maximum for $N=21$ at the starting node
$1$.

At this point it is instructive to look at very small circles of $N=3$ and
$4$ nodes where the analytic results are still handy. With
Eq.(\ref{ampl_bloch2}) we find for the transition probabilities for $N=3$,
\be
\pi_{kj}(t) = \begin{cases} 
\frac{5}{9} + \frac{16}{9}\cos^3 t - \frac{4}{3}\cos t & k=j \\
\frac{2}{9} - \frac{8}{9}\cos^3 t - \frac{2}{3}\cos t &
k\neq j.
\end{cases}
\ee
For $N=4$ we have
\be
\pi_{kj}(t) = \begin{cases} \cos^4 t &
k=j \\
\sin^4 t & k=2j \\
\sin^2 t \cos^2 t & \mbox{else},
\end{cases}
\ee
where $\pi_{jj}(t)$ and $\pi_{j,2j}(t)$ are only shifted by a phase factor
of $\pi/2$ but equal in magnitude.  The limiting probability distributions
are for $N=3$, $\chi_{11} = 5/9$ and $\chi_{12} = \chi_{13} = 2/9$ and for
$N=4$, $\chi_{11} = \chi_{13} = 3/8$ and $\chi_{12} = \chi_{14} = 1/8$,
and thus support the findings for bigger lattices, e.g.\ Fig.\
\ref{1d-circle_chi}.

The occurrence of the regular structures is reminiscent of the so-called
quantum carpets \cite{kinzel1995, fgrossmann1997, berry1996}. These were
found in the interference pattern of a quantum particle, initially
prepared as a gaussian wave packet, moving in a 1d box. The spreading and
self-interference due to reflection of the wave packet at the walls lead
to patterns in the spacetime probability distribution. Furthermore, after
some time, the so-called revival time, the whole initial wavefunction gets
reconstructed. For a particle in a box, theses quantum revivals are
(almost) perfect and the revival time $T$ follows from the energy $E_n =
(n\pi\hbar/L)^2/2m = n^2 2\pi \hbar /T$, where $L$ is the width of the box
\cite{fgrossmann1997}. 

\begin{figure}[ht]
\centerline{
\includegraphics[width=0.425\columnwidth]{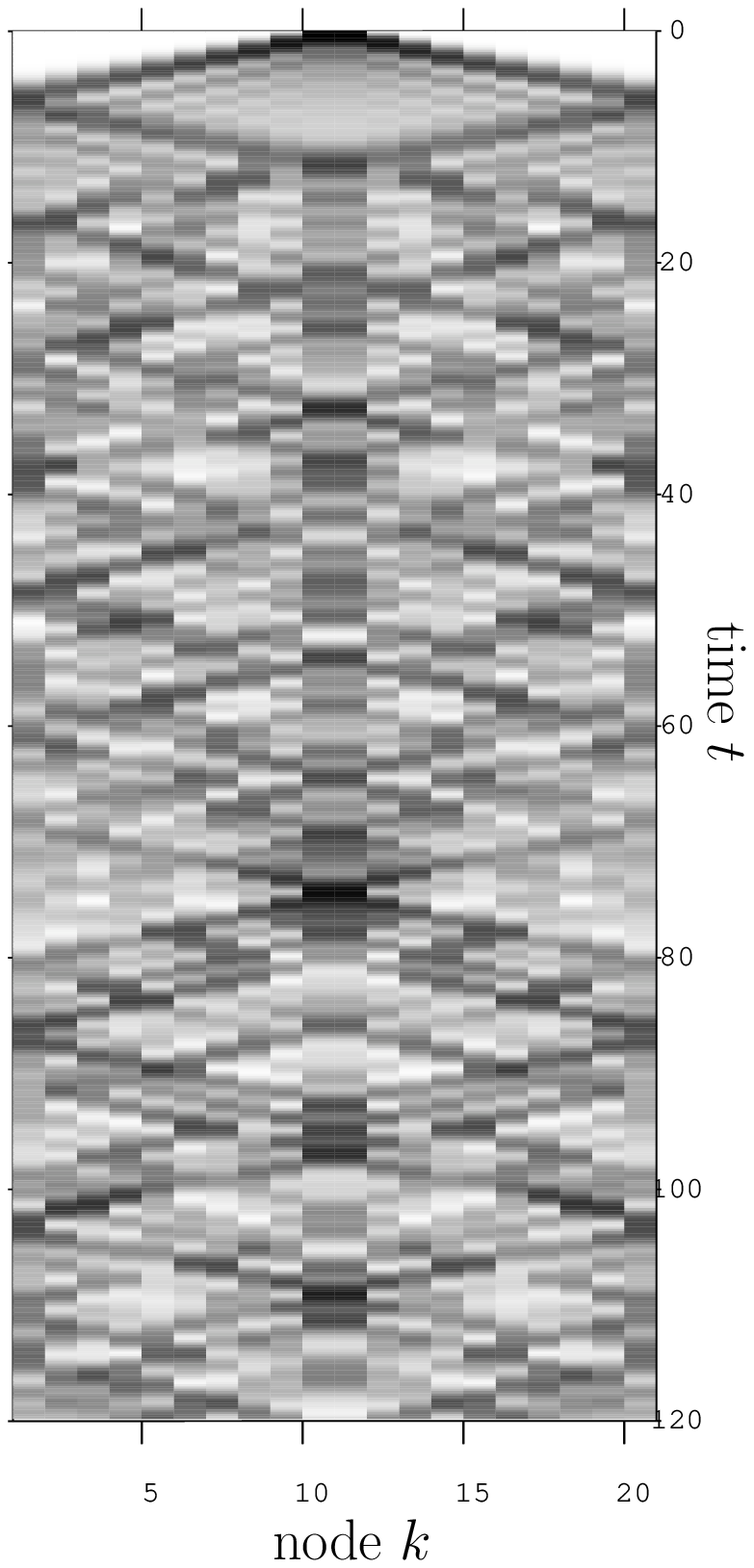}
\includegraphics[width=0.425\columnwidth]{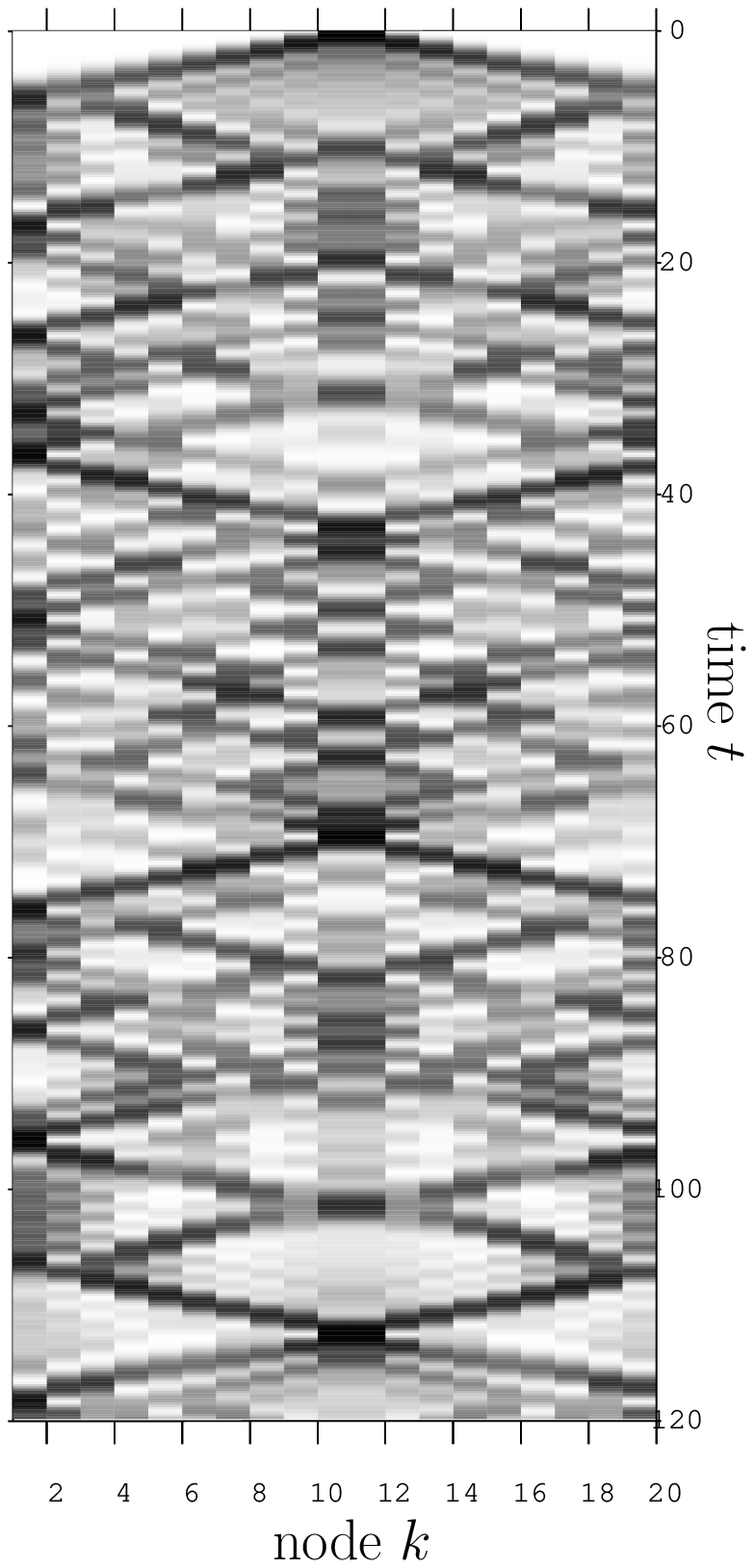}}
\caption{Contour plot of the probability for a CTQW on a circle of length
$N=21$ (left) and $N=20$ (right) over longer times $t$ than in
Fig.~\ref{1d-circle-cont}. Dark regions denote high probabilities. See
Fig.\ref{1d-circle} for units.
}
\label{1d-circle-contlt}
\end{figure}

For very long times, Fig.~\ref{1d-circle-contlt} shows a contour plot of
the probability for a CTQW on a circle of length $N=21$ (left) and $N=20$
(right).  There is obvious structure in the interference pattern.
Furthermore, there are areas on this quantum carpet where there is a very
high probability, visualized by dark regions, to find the CTQW at its
starting point.  Thus, quantum revivals also occur for the discrete
lattice. However, these are not perfect. 

The revival time $\tau$ is given by $\alpha_{kj}(\tau) = \alpha_{kj}(0)$.
Since the transition amplitudes are given as a sum over all modes $n$, see
Eq.(\ref{ampl_bloch2}), we cannot give a universal revival time which is
independent of $n$. Nevertheless, from Eq.(\ref{ampl_bloch2}) we get for
each mode $n$ its revival time
\be
\tau_n=\frac{r\pi}{1-\cos(2n\pi/N)} = \frac{r\pi}{2} [1 + \cot^2(n\pi/N)],  
\label{revtime}
\ee
where $r\in{\mathbb N}$ (without any loss of generality we set $r=1$).
From Eq.(\ref{revtime}) we find that $\tau_n > \tau_{n+1}$ for $n\in
]0,N/2]$ and $\tau_n < \tau_{n+1}$ for $n\in ]N/2,N]$.  For certain values
of $n$, $\tau_n$ will be of order unity, e.g.\ for $n=N/2$ we get $\tau_n
= \pi/2$.  However, for $n << N$, Eq.(\ref{revtime}) yields $\tau_n =
N^2/2\pi n^2 \equiv \tau_0/n^2$, which is analogous to the particle in the
box and where $\tau_0$ is a universal revival time.  Thus, the revival
times $\tau_n$ have large variations in value.  To make a sensible
statement about at least the first revival time, we need to compare it to
the actual time needed by the CTQW for travelling through the lattice. As
mentioned earlier, interference effects in the return probability
$\pi_{1,1}(t)$ are seen after a time $t\approx N/2$. The first revival
time has to be larger than this, because there cannot be any revival
unless the wave reaches its starting node again. Our calculations suggest
that the first revival time will be of order $\tau_0$.  From
Fig.~\ref{1d-circle-contlt} we see that the first (incomplete) revival
occurs for $N=20$ at $t\approx 70 > 20^2/2\pi$ and for $N=21$ at $t\approx
75 > 21^2/2\pi$. 

In conclusion we have shown that CTQWs on regular 1d lattices show regular
structures in their spacetime transition probabilities. By employing the
Bloch function ansatz we calculated quantum mechanical transition
probabilities (as a function of time $t$) between the different nodes of
the lattice.  These results are practically indistinguishable from the
ones obtained by diagonalizing the transfer matrix. We note that the
results obtained via the Bloch function ansatz can be related to recent
results for GCQWs.  The spacetime structures are reminiscent of quantum
carpets, but have their first revival at later times than what is found
for quantum carpets. 

Support from the Deutsche Forschungsgemeinschaft (DFG) and the Fonds der
Chemischen Industrie is gratefully acknowledged.

\end{document}